# On the relationship between rheology and percolation in the gelation of weakly attractive colloids: Beyond "snapshot" percolation

Luis A. Pugnaloni[*]
*Procter Department of Food Science, University of Leeds, Leeds LS2 9JT, UK*
*and*
*Instituto de Física de Líquidos y Sistemas Biológicos, Calle 59 Nro. 789, 1900 La Plata, Argentina*
(11 June 2004)

A discussion on the viscoelastic properties of weakly attractive colloids and the relation to a newly developed percolation theory is presented. The need for taking into account the particle–particle bond lifetime in order to compare rheological and percolation estimates of the sol–gel transition is stressed. Recent molecular dynamics simulations and generalised percolation theory that use a bond lifetime criterion in the definition of aggregates show that the percolation transition can be reconciled with the rheological sol–gel transition. The percolation concept is used here as a "measuring" tool to assess the viscoelastic behaviour of the system.

The sol–gel transition in colloidal systems is characterised by a change in viscoelastic behaviour: form liquid-like (sol) to solid-like (gel) [1]. This transition, however, is not discontinuous. Moreover, whether a viscous or an elastic behaviour is measured depends on the relaxation time scale assessed by the experimental test. This time scale is related to the oscillation frequency (if measuring complex shear modulus) or the shear-rate (for the shear viscosity). This paper focuses on weakly attractive colloids, which present thermoreversible gelation and need relatively high densities to gel. These types of systems are amenable of study under the equilibrium statistical mechanics theory.

Percolation theory is aimed to describe the transition between a state where the system consists of a set of finite size aggregates to a state where an infinite spanning cluster connects the edges of the system [2]. Several applications of this concept to colloids have been described such as the insulator–conductor transition in water-in-oil microemulsions [3] and the abovementioned sol–gel transition [4].

The classical picture that relates percolation with the sol–gel transition is exemplified in Fig. 1. Starting from a sol state, an increase in the volume fraction ($\phi$) of the particles leads to the formation of ever growing aggregates due to weak attractive particle–particle interactions. Eventually, one or more aggregates span the system. This spanning cluster(s) should be able to transmit stresses between opposite sides of the system. In doing so, the system would present a sharp increase in the complex shear modulus and would have undergone a transition from sol to gel. As more and more particles are incorporated to the spanning aggregate by a further increase in $\phi$, the elastic behaviour of the system builds up due to the increasing strength of the particle network.

This interpretation of relation between structural changes (percolation) and rheological changes (gelation) in colloidal systems has been used for decades. However, this classical picture does not take into account the fact that the same colloidal sample can present either liquid-like or solid-like behaviour depending on the deformation-rate impose by the rheometer. Generally, at high deformation-rates the system tends to behave like a solid (high elastic modulus, low loss modulus and shear thinning response), whereas at low deformation-rates the system behaves like a liquid (low elastic modulus and Newtonian response). This deformation-rate effect is essential to understand the microscopic dynamics of the colloid. Any percolation theory aimed to describe the sol–gel transition has to be able to account for this effect. Otherwise, any agreement or disagreement between rheology and percolation can be overturned be simply using a different deformation-rate in the rheometer. This paper shows how a generalised percolation theory is able to describe the deformation-rate effect in the sol–gel transition.

A schematic plot of the shear storage (elastic) modulus, $G'$, and shear loss (viscous) modulus, $G''$, as a function of the frequency, $\omega$, is presented in Fig. 2a. In Fig. 2b, the corresponding shear viscosity, $\eta$, of the system is shown as a function of the shear-rate, $\dot\gamma$. These general trends are followed by many weakly attractive colloids [5] at moderate $\phi$ (roughly 0.15–0.45). The characteristic frequency, $\omega_c$, at which $G' = G''$ and the characteristic shear-rate, $\dot\gamma_c$, at which the change from Newtonian (constant viscosity) to shear thinning are associated to a change from liquid-like to solid-like behaviour. Both quantities give roughly the same estimate for the characteristic relaxation time, $\tau_c \approx \omega_c^{-1} \approx \dot\gamma_c^{-1}$, of the system. Clearly, if a sol–gel transition line has to be defined on the temperature–volume fraction ($T-\phi$) phase space by using rheology, the working frequency, $\omega$ (or the working shear-rate, $\dot\gamma$) of the rheometer will have a great effect on the position of this transition line.

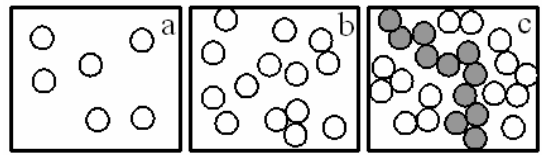

FIG. 1. The classical snapshot percolation transition. From (a) to (c), an increase in the volume fraction promotes aggregation of the particles. In (c), the first spanning (grey) cluster appears.

Indeed, working at very high frequencies (or very high shear-rates) will result in a sol–gel transition line at relatively low $\phi$, since a solid-like response will be expected for $\omega \gg \omega_c$. Conversely, working at very low frequencies (or very low shear-rates) will result in a sol–gel transition line at relatively high $\phi$, since the system will have time to relax for $\omega \ll \omega_c$. A proper comparison between the rheological sol–gel transition and its percolation transition analogue has remained elusive for weakly attractive colloids so far; probably due to this deformation-rate dependence of the sol–gel transition line that cannot be accounted for in the classical "snapshot" percolation theory.

Computer simulation and connectedness theory to study the percolation —*i.e.* the appearance of a spanning cluster— for continuum systems have been used systematically since the paper by Coniglio and co-workers [6]. The most noticeable system on which percolation studies [7] have concentrated in relation to colloids is the Baxter's sticky spheres model [8]. A probably more realistic system investigated in the percolation literature is the square well system [9]. Generally, in Monte Carlo (MC) and molecular dynamics (MD) simulations, a system is said to be in a percolated state if 50% of the configurations (in MC) or 50% of the time (in MD) a snapshot of the system shows a spanning cluster (or aggregate). A cluster is defined as a set of interconnected particles. This requires a connectivity criterion to decide whether two specific particles are directly connected (bonded) or not. For sticky spheres the criterion is "to be in contact"; for the square well system the connectivity criterion can be either, that the particles are within the range of the potential well, or that the relative potential energy outweighs the relative kinetic energy of the pair of particles. With more realistic continuum pair interaction this last criterion is preferred [10] although many studies use a simple connectivity distance [11].

Since the classical snapshot percolation considers only the configuration of the system at any given time, any spanning aggregate detected is not guaranteed to last for any minimum period of time. In particular, at the percolation transition one expect that the spanning cluster will last for only an infinitesimal time. For a spanning cluster to transmit stresses across the system under shear, a minimum lifetime is necessary for the cluster to stretch under the externally imposed deformation. Of course, the more slowly the deformation is applied, the more long-lasting the cluster has to be in order to stretch and transmit the stress before thermal agitation breaks it up into disconnected sub clusters. Therefore, the percolation transition described by the classical snapshot percolation corresponds only to the infinite frequency (or infinite shear-rate) rheological sol–gel transition, for which the assessed time scales are far shorter than the characteristic relaxation time, $\tau_c$, of the system.

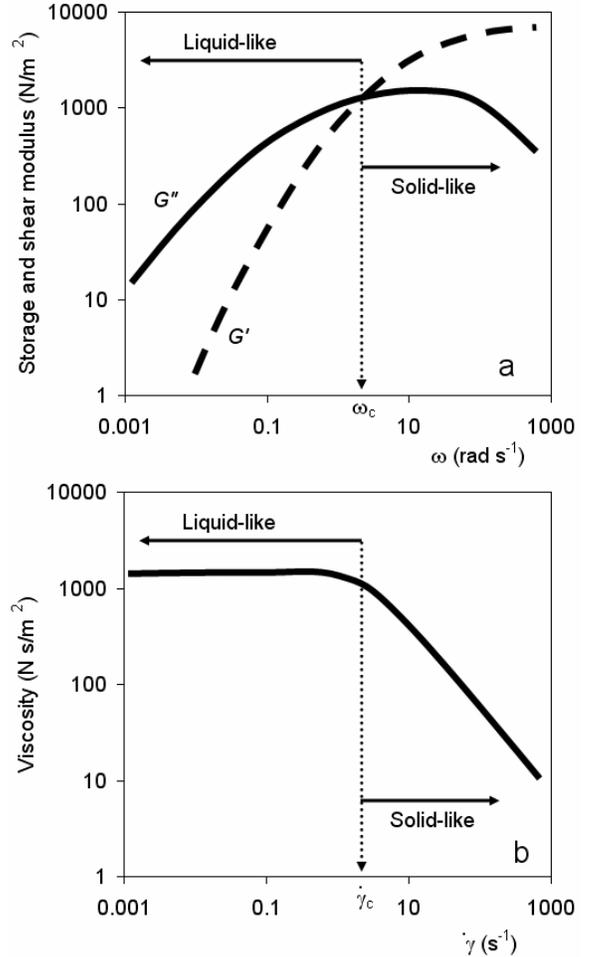

FIG. 2. (a) Storage, $G'$, and loss, $G''$, moduli as a function of the oscillation frequency, $\omega$. (b) Shear viscosity, $\eta$, as a function of the shear-rate, $\dot{\gamma}$. The liquid-like and the solid-like regimes are indicated along with the characteristic frecuency, $\omega_c$, and characteristic shear-rate, $\dot{\gamma}_c$.

It can be argued that, at the percolation transition, as a spanning cluster breaks up due to thermal agitation another cluster is formed to span the system. As a consequence, a spanning cluster is always present in the system to transmit stress. However, any newly formed spanning cluster has to deform before it can transmit stress and at a slow deformation-rate it will break up before it can be of any effect.

Only recently [12,13], continuum percolation theory has been extended to account for the lifetime of the particle–particle bonds in the connectivity criterion. First of all, it has to be emphasized that the cluster size distribution —and the percolation transition in particular— is dependent on the connectivity criterion used to identify bonded pairs [14]. This does not mean that the system changes its behaviour —which is determined by the corresponding partition function—; rather, it is the cluster detection mechanism that changes.

Therefore, the choice of a particular connectivity criterion reflects the type of clusters in the system that one is trying to pick up. This also applies to any rheology experiment: the shear frequency (or the shear-rate) determines the timescales assessed and hence the type of structures that can make an impact on the shear modulus (or the shear viscosity).

In Ref [12] the authors introduce a new connectivity criterion according to which two particles are said to be bonded if they remain within a certain distance, $d$, during a minimum period of time, $\tau$. Molecular dynamics results for a Lennard–Jones system and a generalized integral equation theory to deal with this more complex connectivity criterion have been reported [13].

The new criterion is able to distinguish between aggregates having different lifetimes. By setting $\tau$ to a non-zero value one can ensure that detected bonds between any pair of particles last for at least $\tau$ seconds. Therefore, any cluster detected will last for at least $\tau$ seconds. This means that a spanning cluster of this type will survive over $\tau$ seconds and will be able to transmit stresses when sheared at finite (as oppose to infinite) $\omega$ (or $\dot{\gamma}$). Moreover, the rheological sol–gel transition line drawn by using a finite $\omega$ (or $\dot{\gamma}$) value should coincide with the percolation line drawn by using an associated non-zero $\tau$ value.

In Fig. 3, the percolation line for a Lennard–Jones system [13] for two values of $\tau$ (0 and 0.5 in reduced units) obtained by MD is shown. As expected, a non-zero value of $\tau$ shifts the percolation line to higher densities. This is in qualitative agreement with the rheological sol–gel transition line drawn at infinite $\omega$ (or $\dot{\gamma}$) (i.e. $\tau = 0$) and at a finite $\omega$ (or $\dot{\gamma}$) (i.e. $\tau > 0$). Recently, Zarragoicoechea et. al. [15] have obtain excellent agreement between this MD results and a generalized connectivity integral equation theory. Of course, a Lennard–Jones system is not appropriate to describe the behaviour of a colloidal system. However, this trend will be encountered for any other interacting potential. The application of the new percolation theory to new models that include activation barriers for the bond breakage in continuum [16] and lattice [17] systems would be of particular interest.

As mentioned above, the infinite-$\omega$ (or $\dot{\gamma}$) limit corresponds to the classical "snapshot" percolation ($\tau = 0$). This zero-$\tau$ percolation line defines the minimum $\phi$ required, at given $T$, to find at least one ephemeral spanning cluster. It would be interesting to be able to associate a physical meaning also to the zero-$\omega$ (or $\dot{\gamma}$) limit (i.e. the infinite-$\tau$ limit). One is tempted to infer that this limit constitutes the transition to a glass state. At $\tau \to \infty$ ($\omega = 0$), only a state where the system relaxes infinitely slowly will be perceived as solid-like. However, an exhaustive investigation on the percolation line for $\tau \to \infty$ has to be carried out before any conclusion can be drawn.

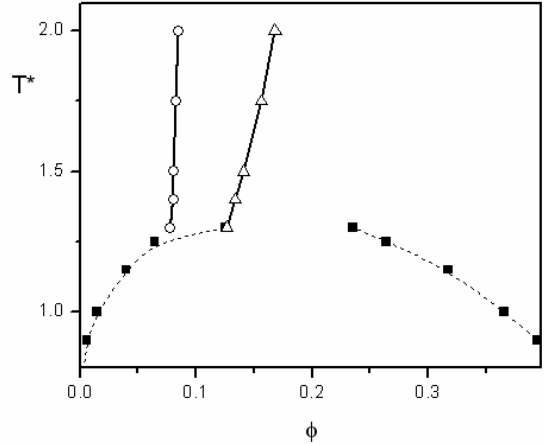

FIG. 3. Temperature–volume fraction phase diagram for a Lennard–Jones system. Dotted line and full squares correspond to the gas–liquid coexistence line [18]. Solid line and open symbols correspond to the percolation line for $\tau = 0$ (circles) and $\tau = 0.5\sigma\sqrt{m/\varepsilon}$ (triangles) [13]. The connectivity distance is $d = 1.5\sigma$. Here, $m$, $\varepsilon$ and $\sigma$ are the mass of the particles, the energy parameter in the Lennard–Jones potential and the diameter of the particles, respectively. Temperature is in reduced units: $T^* = k_B T / \varepsilon$

In summary, the introduction of a bond lifetime prerequisite in the connectivity criterion in the percolation theory allows to account for the deformation-rate effect in the rheological sol–gel transition line on the $T-\phi$ phase diagram. Since experiments on gelation are generally performed at finite working frequency (or shear rate) the classical "snapshot" percolation line ($\tau = 0$, $\omega \to \infty$) will always be at the left (lower $\phi$) of the experimental results (see for example figure 13 in Ref. [19]). The new connectivity criterion will certainly shift the percolation line towards the rheological sol–gel transition line as soon as a non-zero value for $\tau$ is introduced. The exact mathematical relation between $\omega$ (or $\dot{\gamma}$) and $\tau$ remains to be defined. However, recent molecular dynamics simulations [20] of the shear of Lennard–Jones particles suggest that, for this system at least, $\tau = 0.07\dot{\gamma}^{-1}$. At present we are working in the simulation of the rheology and percolation properties of colloidal systems using existing Brownian dynamics models [21] to shed light on this relation.

As a final comment, it is worth mentioning that many microscopy studies on gels have failed to correlate the gel structure with the rheology of the system. From the previous discussion it can be argue that microscopy suffers from the same drawback as the classical snapshot percolation theory, namely, a lack of control over the timescales assessed. It would be interesting to investigate the structures seen under the microscope if images are taken with different exposure times. One can expect that long exposure time images will correlate with low deformation-rate rheology, whereas short exposure time images will correlate with high deformation-rate rheology.

I am grateful to Prof. Fernando Vericat, Dr. Rammile Ettelaie, Prof. Eric Dickinson, and Dr. Frank Podd for fruitful discussions on this paper. I am also grateful to Dr. Guillermo Zarragoicoechea who sheared his numerical results prior to publication. The author is a member of CONICET (Argentina).

* E-mail: l.a.pugnaloni@food.leeds.ac.uk; luis@iflysib.unlp.edu.ar

[1] R. G. Larson, *The Structure and Rheology of Complex Pluids* (Oxford University Press, New York, 1999); V. Trappe and P. Sandkühler, Current Opinion Colloid Int. Sci. **8**, 494 (2004).
[2] D. Stauffer and A. Aharony, *Introduction to Percolation Theory* (Taylor and Francis, London 1992).
[3] S. H. Chen *et. al.*, J. Phys.: Condens. Matter **6**, 10855 (1994).
[4] See for example F. Mallamace *et. al.*, Physica A **302**, 202 (2001).
[5] See for example Ref [1] pages 16 and 18; also T. Annable *et. al.*, J. Rheology **37**, 695 (1993).
[6] A. Coniglio *et. al.*, J. Phys. A **10**, 219 (177) ; A. Coniglio, U. De Angelis and A. Forlani, *ibid.* **10**, 1123 (1977).
[7] Y. C. Chiew and E. D. Glandt, J. Phys. A **16**, 2599 (1983); M. A. Miller and D. Frenkel, arXiv:cond-mat/0404318.
[8] R. J. Baxter, J. Chem. Phys. **49**, 2770 (1968).
[9] S. C. Netemeyer and E. D. Glandt, J. Chem. Phys. **85**, 6054 (1986); Y. C. Chiew and Y. H. Wang, J. Chem. Phys. **89**, 6385 (1988).
[10] L. A. Pugnaloni, I. F. Márquez and F. Vericat, Physica A **321**, 398 (2003).
[11] F. Bresme and J. L. F. Abascal, J. Chem. Phys. **99**, 9037 (1993); D. M. Heyes and J. R. Melrose, Mol. Phys **66**, 1057 (1989); C. M. Carlevaro, C. Stoico and F. Vericat, J. Phys.: Condens. Matter **8**, 1857 (1996).
[12] L. A. Pugnaloni and F. Vericat, Phys.. Rev. E. **61**, R6066 (2000).
[13] L. A. Pugnaloni and F. Vericat, J. Chem. Phys. **116**, 1097 (2002).
[14] T. L. Hill, J. Chem. Phys. **23**, 617 (1955).
[15] G. Zarragoicoechea et. al. (to be published).
[16] I. Saika-Voivod *et. al.*, arXiv:cond-mat/0403320.
[17] E. Del Gado *et. al.*, Phys. Rev. E **69**, 051103 (2004).
[18] A. Z. Panagiotopoulos, Mol. Phys. **61**, 813 (1987).
[19] F. Laflèche, D. Durand and T. Nicolai, Macromolecules **36**, 1331 (2003).
[20] L. Angelani, *et. al.*, Phys. Rev. E **66**, 061505 (2002)
[21] M. Whittle, E. Dickinson, Mol. Phys. **90**, 739 (1997).